# Bubble Nucleation and Growth on Microstructure Surface under Microgravity

Qiushi Zhang[1], Dongchuan Mo[1], Jiya Janowitz[2], Dan Ringle[2], David Mays[2], Andrew Diddle[2], Jason Rexroat[2], Eungkyu Lee[1,*], and Tengfei Luo[1,3,*]

1. Department of Aerospace and Mechanical Engineering, University of Notre Dame, IN, USA

2. Space Tango Inc., 611 Winchester Rd. Lexington, KY, USA

3. Department of Chemical and Biomolecular Engineering, University of Notre Dame, IN, USA

* Corresponding authors: eleest@khu.ac.kr; tluo@nd.edu

# ABSTRACT

Understanding the nucleation and growth dynamics of the surface bubbles generated on a heated surface can benefit a wide range of modern technologies, such as the cooling systems of electronics, refrigeration cycles, nuclear reactors and metal industries, etc. Usually, these studies are conducted in the terrestrial environment. As space exploration and economy expanding at an unprecedented pace, the aforementioned applications that potentially deployable in space call for the understanding of thermal bubble phenomena in a microgravity setting. In this work, we investigate the nucleation and growth of surface bubble in space, where the gravity effect is negligible compared to the earth. We observe much faster bubble nucleation, and the growth rate can be ~30 times higher than that on the earth. Our finite element thermofluidic simulations show that the thermal convective flow due to gravity around the nucleation site is the key factor that effectively dissipates the heat from heating substrate to the bulk liquid and slows down the bubble nucleation and growth processes. Due to the microgravity field in space, the thermal convective flow is negligible compared to the terrestrial environment, leading to the localization of heat around the nucleation site, and thus enables faster bubble nucleation and growth in space. We also find that bubble nucleation can be influenced by the characteristic length of the microstructures on the heating surface. The microstructures behave as fins to enhance the cooling of the surface. With finer microstructures enabling more efficient surface to liquid heat transfer, the bubble nucleation takes longer.

**INTRODUCTION**

Since the first expression of inertially controlled growth and collapse of vapor bubble was developed by Lord Rayleigh in 1917,[1] the dynamics of surface bubble have been extensively studied both experimentally and theoretically.[2–10] Understanding the dynamics of surface bubble nucleation and growth can help to formulate the heat transfer models in a wide range of modern technologies, such as the cooling of electronics, refrigeration cycles, nuclear reactors and metal industries, etc.[11] In addition to the traditional pool boiling, surface bubbles can also be generated through the photo-thermal evaporation process driven by enhanced surface plasma resonance heating effect, and the corresponding mechanisms and applications (e.g., particle deposition and sensing) have been investigated in recent decades.[12–18]

Although extensive research has been done to study the dynamics of surface bubble nucleation and growth in a variety of conditions and settings, most of these works were conducted in the terrestrial gravity environment. As we know, surface bubble nucleation and growth are initiated and dictated by the heat transfer between the heating surface and surrounding liquid, and the heating surface temperature can be significantly influenced by the liquid flow close to the bubble nucleation site.[19–21] Therefore, the bubble dynamics in microgravity environment, i.e., in space, can be very different compared to those on the earth because of the distinct heat transfer efficiency and pattern from the heated surface to the surrounding liquid.[22–29] Recently, some primary experimental observations of surface bubble generation in pool boiling in space were reported by Ronshin et al.[30] They measured the geometries of the surface bubbles in space, and observed the non-linear bubble volume growth, which is different from the linear bubble volume growth observed on the earth.[12] The Marangoni flow around the surface bubble and its influence on the boiling heat transfer in space were investigated in Refs.[31,32] The authors found that the Marangoni effect was more significant and the flow pattern was different in space that changed the temperature profile around the bubble, which resulted in a higher bubble growth rate. In addition to bubble growth, the collapse, detachment, coalescence and dispersion of bubbles in liquid under microgravity were also studied in previous works.[33,34]

The nucleation and growth of surface bubble involve a complex interplay of physical phenomena, and a comprehensive understanding of these phenomena requires consideration of multiple disciplines, including mass transfer, gas diffusion, fluid mechanics, thermodynamics, etc.[35–37] Therefore, precisely predicting the overall dynamics of a surface bubble using finite element method can still be very challenging given the limitations in the accuracies of the model geometries, mesh density and time step size, as well as the approximations in the physical properties of fluid that we usually employed in finite element simulations.[23,38–40] On the other hand, despite the insights gained from studying bubble dynamics in space can benefit many important practical applications, experimental studying surface bubble nucleation and growth dynamics in space is still uncommon due to the technical challenges in conducting experiments in the unique environmental conditions associated with the high experimental costs.[41–43] One of the major fluid flows occurs in pool boiling heat transfer that changes dramatically from terrestrial gravity to microgravity environment is thermal convective flow.[44–48] Thermal convective flow is produced by the temperature-gradient induced density gradient in a fluid. The hotter fluid with lower density rises upwards while the colder fluid moves downwards driving by the buoyancy force on the earth. However, the buoyancy force in space is almost negligible due to the microgravity environment, largely reducing the significance of thermal convection flow effect. Owing to the difficulties in experimental approaches, the detailed analysis of how the reduced thermal convection flow influences surface bubble nucleation and growth dynamics in space, and their comparisons to the terrestrial experiments are still lacking.

In this work, we carried out experiments onboard the international space station (ISS) to study the nucleation and growth of surface bubbles on heated substrates with different microstructures under microgravity. Videography revealed that surface bubbles nucleated and grew much faster in space than those on the earth. Our thermofluidic finite element simulations attributed the unique bubble dynamics in space to the effects of the reduced thermal convective flow. Additionally, we also studied the influence of the characteristic length of surface microstructures on bubble nucleation. Bubble dynamics on nano/micro-structure pre-decorated surfaces have also attracted many research attentions in recent decades.[49–52] For instance, Liu et al.[53] and Chen et al.[54] found the densities and geometries of the gold nanopillars and micropyramids on surfaces can significantly influence the collective input heating power, and thus affect the nucleation time of surface bubble. Dong et al.[55] found that both the characteristic length and wettability of surface microstructures can affect the dynamics of surface bubble. When the characteristic length of a surface microstructure is in the range of 5 ~ 100 times less than bubble radius, the micro curvature can significantly influence the bubble dynamics, otherwise the wettability effect predominates. In our experiments, the substrates with different porosities but similar wettability have the microstructure characteristic lengths in the range of about 100 ~ 500 nm while a stabilized growing bubble (after nucleation process) usually has a radius in millimeters scale. Therefore, our study on the influence of microstructure characteristic length mainly focused on the bubble nucleation process while the radius of the bubble is still in micrometer scale. We found the microstructures function as fins to enhance the cooling of the surface. With finer microstructures enabling better surface to liquid heat transfer, the bubble nucleation takes longer. These results revealed interesting physics and may push the boundaries of the knowledge in this field to benefit many space and terrestrial applications, such as phase change cooling and sensing.[56–58]

**RESULTS AND DISCUSSION**

Here, a series of Cu microstructured substrates were used to conduct heat into the boiling system, which were fabricated by the so-called hydrogen bubble template electrodeposition method (**Figure 1a** and Supporting Information SI1).[59–62] A DC power supply was applied to the Cu cathode and anode substrates that immerged in the $H_2SO_4/CuSO_4$ solution. Due to the external electric field applied, $Cu^{2+}$ ions in the solution moved toward and were finally deposited onto the Cu cathode substrate that would be used to generate surface bubbles in the pool boiling experiments later. The molarity of $H_2SO_4$ in the $H_2SO_4/CuSO_4$ solution was fixed as 0.8 M, and by controlling the molarity of $CuSO_4$ while keeping the electric current applied (1.0 A·cm$^{-2}$) and deposition time (60 s) as the same, we can control the porosity of the microstructures on Cu substrates, leading to different structure characteristic lengths. There are four microstructured Cu substrates with different characteristic lengths that had been fabricated and investigated in this project, which were labeled as C1 with 0.2 M, C2 with 0.4 M, C3 with 0.8 M, and C4 with 1.0 M molarity of $CuSO_4$ (**Figure 1b**). As the optical images shown in **Figure 1b**, the characteristic length of the microstructure increases as the molarity of $CuSO_4$ increases.

A fabricated Cu substrate was then attached with thermal epoxy onto the inner wall of a quartz cuvette with the internal dimensions of 10 mm (H) × 20 mm (W) × 43.75 mm (L) and a wall thickness of 1.25 mm (**Figure 1c**). A Peltier with a 10 mm × 10 mm surface area was affixed to the outside of the cuvette such that the heat was conducted through the quartz, epoxy, and eventually to the Cu substrate for the surface bubble nucleation to occur. We note the thickness of the epoxy is much thinner than the thickness of the quartz cuvette wall. The Cu substrate was trimmed to fit the inner width of the cuvette, so they are slightly less than 20 mm. The imaging process is also depicted in **Figure 1c**. The imaging axis of the camera was aligned with a small angle of ~10 degrees to the substrate plane, and a LED background light was used as the light source. All videos were captured at 110 FPS and 2 megapixels resolution. We first used the camera to image a grid with an inter-line distance of 1 mm, which was then used as the pixel-to-real-size converter to estimate the real sizes of the surface bubbles in the videos. To note, the bubble was generated on the Cu substrate that attached on the top inner wall of the cuvette while the gravity is downward in the terrestrial experiments (see **Figure 1c**). This setup can accelerate the bubble generation process and prevent its detachment from the surface. Then, the whole setup was facilitated into an integrated instrument box, named 'CubeLab', by Space Tango (**Figure 1d**). One thing to note is that all the experiments in space were conducted inside the NASA International Space Station, which means the experimental condition was ambient pressure rather than vacuum. The CubeLab was launched to the ISS twice via SpaceX Cargo Dragon 22 and Northrop Grumman, respectively. The experimental processes were monitored on the earth, and the recorded videos were downlinked for detailed analysis. The terrestrial experiments were performed in the CubeLab prior to the launch to ensure that the only difference between the sets of experiments was from gravity.

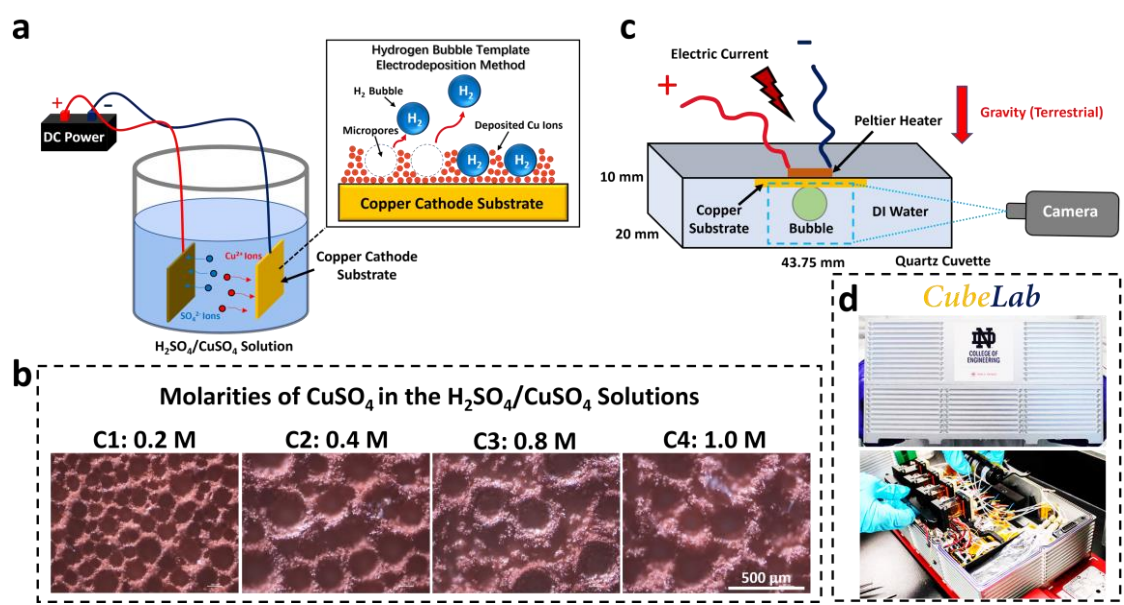

**Figure 1.** (a) The schematic of the hydrogen bubble template electrodeposition method used to fabricate the microstructured Cu substrates. (b) The optical images showing the Cu substrates (C1 ~ C4) with different porosities, using the molarities of $CuSO_4$ 0.2 ~ 1.0 M, respectively. (c) The schematic of the setup to generate surface bubble by surface heating and monitor its nucleation and growth processes. (d) The integrated instrument, 'CubeLab', developed by Space Tango for this project.

**Figures 2a and b** show several typical frames from the recorded videos in the terrestrial and space conditions (also see Supporting Movies S1 and S2). The upper, middle and lower panels are the moments of surface bubble nucleation, growth and the final phase at the end of the video, respectively. In **Figures 2a and b**, the heating power, cuvette setup, volume and air concentration of DI water were all kept the same in the experiments on the earth and in space. The C4 substrate with the largest characteristic length (see **Figure 1b**) was used in both cases. Therefore, the only difference between the experiments in the terrestrial and space conditions is if there was gravity influencing the fluid flow during the bubble formation or not. Comparing the snapshots of the experiments in the terrestrial (**Figure 2a**) and space (**Figure 2b**) conditions, we first found that the nucleation of space bubble was obviously faster than terrestrial bubble (upper two panels). The bubble nucleation occurred at around 76 s in space after we started heating, while nucleation took about twice of heating time and started at 161 s in the terrestrial condition with the same experimental setup. Besides, as we can see in the middle two panels, the space bubbles were much larger than the terrestrial bubbles at the same time (150 s) after nucleation, which means the growth of space bubble was also much faster. Finally, it is interesting that the space bubbles suddenly collapsed after the heating process lasted for a certain period of time (213 s), but the terrestrial bubbles never reached that phase throughout the whole heating process that lasted for ~600 s (lower two panels).

To further quantify the difference in surface bubble growth, we plotted the volumes of space and terrestrial bubbles as a function of time after nucleation. As shown in **Figure 2c**, the volume of terrestrial bubble (black) grows almost linearly with time, which is consistent with previous findings.[12,13] However, the volume of space bubble (red) grows much faster, and the size can reach about 10 ~ 20 times larger than terrestrial bubble. It is interesting that the volume growth of space bubble is nonlinear with time, suggesting a different surface bubble growth mechanism in space, and we will further investigate it in the following section (**Figure 5**). In **Figure 2d**, we plotted the volume growth rates of space and terrestrial bubbles with time in the log scale. The volume growth rate of terrestrial bubble is relatively stable during most of the growth stage, but the growth rate of space bubble has increased by ~2 orders of magnitude during the same period, and it finally reaches ~30 times greater than the volume growth rate of terrestrial bubble before collapsing.

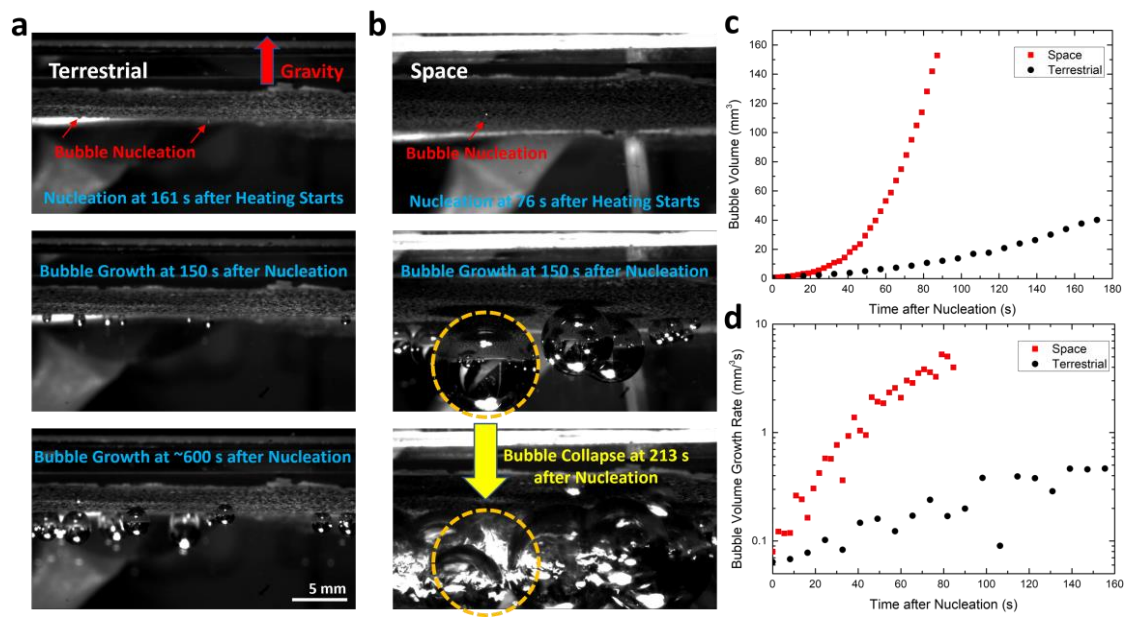

**Figure 2.** The snapshots showing the moments of surface bubble nucleation (upper), growth (middle) and the final phase (lower) at the end of the video in terrestrial (a) or space (b) condition. (c) The volumes of space (red) and terrestrial (black) bubbles as a function of time after nucleation. (d) The volume growth rates of space (red) and terrestrial (black) bubbles as a function of time after nucleation in log scale.

In order to understand the experimental findings and compare the different surface bubble dynamics from the earth to space, we performed thermofluidic simulations using finite element method to help analyze the bubble nucleation and growth processes. The model used to simulate the nucleation of space and terrestrial bubbles is shown in **Figure 3a** with more details of the model setup and simulations included in the Supporting Information (SI2). The flow effect, thermal conduction and thermal convection were included in our transient model, and all the geometries were built according to the real dimensions of experimental setup. In this 2D model, a large box of water (60 mm × 20 mm) is sandwiched by two thin layers of solid $SiO_2$ (60 mm × 1 mm). A thin layer of microstructured Cu substrate (5 mm × 0.2 mm) is immerged in the water, which is the heating source of the boiling system. The geometry of the microstructure on Cu substrate was built according to the characteristic length of the C4 substrate (**Figure 1b**). By switching on or off the gravity effect in these simulations, we can mimic the terrestrial and space conditions, respectively.

A simulated temperature profile of the Cu substrate is shown in the insert of **Figure 3a**. Temperature is distributed symmetrically along the horizontal axis, while the maximum surface temperature is located at the center of substrate. Surface bubble nucleation usually starts when the surface temperature of heating substrate reaches the nucleation temperature, and the nucleation temperature of gas-saturated DI water at ambient pressure is reported to be ~422 K.[18,63] Therefore, we plotted the maximum substrate surface temperature as a function of heating time, as shown in **Figure 3b**. The maximum substrate surface temperature increases much faster in space (red), and it can reach ~50 K higher than the terrestrial case after the heating process lasts for ~15 s. Then, we added the nucleation temperature line (blue dash line), 422 K, into the plot, and found that the terrestrial model needs about twice of heating time in order to reach the nucleation temperature, i.e., nucleation time, compared to space model. Although the amplitudes of the nucleation times in the simulations are different from the real experiments due to some limitations in model geometries (e.g., exact surface morphology), and the fact that a 2D model was being used to simulate the 3D reality, these simulation results nevertheless reproduce the experimental observations that surface bubble nucleates much faster in space as in **Figures 2a and b** (upper two panels).

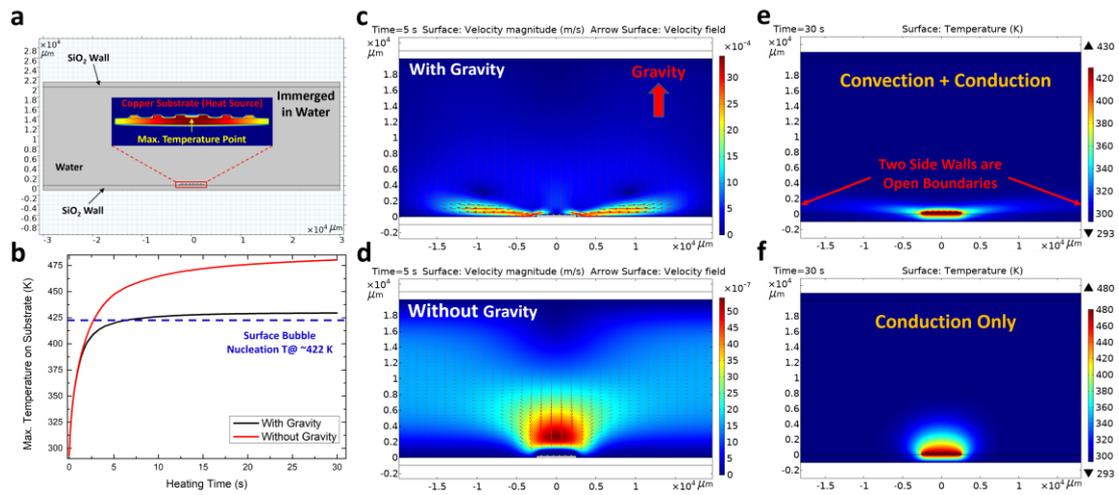

**Figure 3.** (a) The model used in the thermofluidic surface bubble nucleation simulations. (b) The calculated maximum substrate surface temperatures as a function of heating time in the terrestrial (black) and space (red) conditions. The bubble nucleation temperature at ~422 K is indicated by a blue dash line. The simulated fluid velocity fields at t = 5 s in the terrestrial (c) and space (d) conditions. The simulated temperature profiles at t = 30 s in the terrestrial (e) and space (f) conditions.

The absence of gravity is the key reason why substrate surface temperature increases faster in space than that on the earth (**Figures 2c and d**). As we discussed above, the density gradient induced by temperature gradient can lead to thermal convection in gravity field. This is evidenced by the simulated fluid velocity field in the terrestrial model shown in **Figure 3c**. Strong circulation is formed on each side of the heated substrate with opposite directions, and the magnitude of flow velocity can be as high as ~$10^{-3}$ m/s. The feature of fluid circulation indicates that there is significant thermal convection flow in the liquid.[64] However, due to the absence of gravity, thermal convection does not contribute to the fluid flow field in space model (**Figure 3d**), leading to the flow velocity dropping by ~3 orders of magnitude to ~$10^{-6}$ m/s. The weak fluid flow in space is only due to the expansion of the hotter liquid near the heating substrate.[65] The flow field can influence the temperature profile in the boiling system. In the terrestrial model, the thermal convective circulation will grow increasingly larger during the heating process, transferring heat away from the hot substrate to the bulk liquid (**Figure 3e**). This makes the substrate surface temperature in the terrestrial model to increase slower than the in-space counterpart, where heat transfer is dominated by conduction, and thus heat is more localized around the substrate surface, leading to faster surface temperature rise and hence earlier bubble nucleation (**Figure 3f**).

We also studied the influence of surface microstructure on bubble nucleation time. As shown in **Figure 1b**, we prepared four microstructured substrates with a range of characteristic lengths (100 ~ 500 nm). We conducted boiling experiment using each of these substrates in space while all the other experimental parameters and setup were kept the same. The heating power was tuned down in this set of experiments compared to **Figure 2** in order to magnify the difference in nucleation times among these substrates. As shown in **Figure 4a**, the nucleation time decreases monotonically as the characteristic length increases (also see Supporting Movies S3 and S4). That is to say, the finer structure usually requires longer time to nucleate. To understand this, we repeated the thermofluidic simulations to compare the surface temperature profiles of the finest (C1) and the coarsest substrate (C4). The microstructures were modeled as walls standing on the substrates with the spacing set as the average characteristic length of the micropores obtained from the experimental characterization of the corresponding substrates (**Figure 1b**). In the simulations, the heat generation rate of the two substrates was kept the same, and no gravity was considered. **Figure 4b** shows that the maximum substrate surface temperature increases as a function of heating time on each of the two substrates. Although the temperature difference between the two substrates is not as large as that between the terrestrial and space models in **Figure 2**, we still can find the C1 substrate needs slightly longer time to reach the nucleation temperature than C4 substrate. The simulated temperature profiles around the heating substrates are shown in **Figure 4c**. Since both substrates were simulated in the space setting, the heat from substrate can only be dissipated by conduction. Those microstructures on substrate surfaces can behave as fins to enhance the heat conduction – an effect seem extensively for convective interfaces, but also observed in conductive interfaces.[66] Comparing to the coarser surface, the finer surface has a denser fin structure, resulting in better heat conduction across the interface,[67,68] which helps cool the surface more efficiently than the coarser surface. We also note bubble nucleation can also depends on the nucleation site and the trapped gas in microstructures.[69–71] However, it is expected that the finer structures can provide more nucleation sites and more easily trap gas than the coarser structures, which thus should not be the root cause for the observed trend in nucleation times.

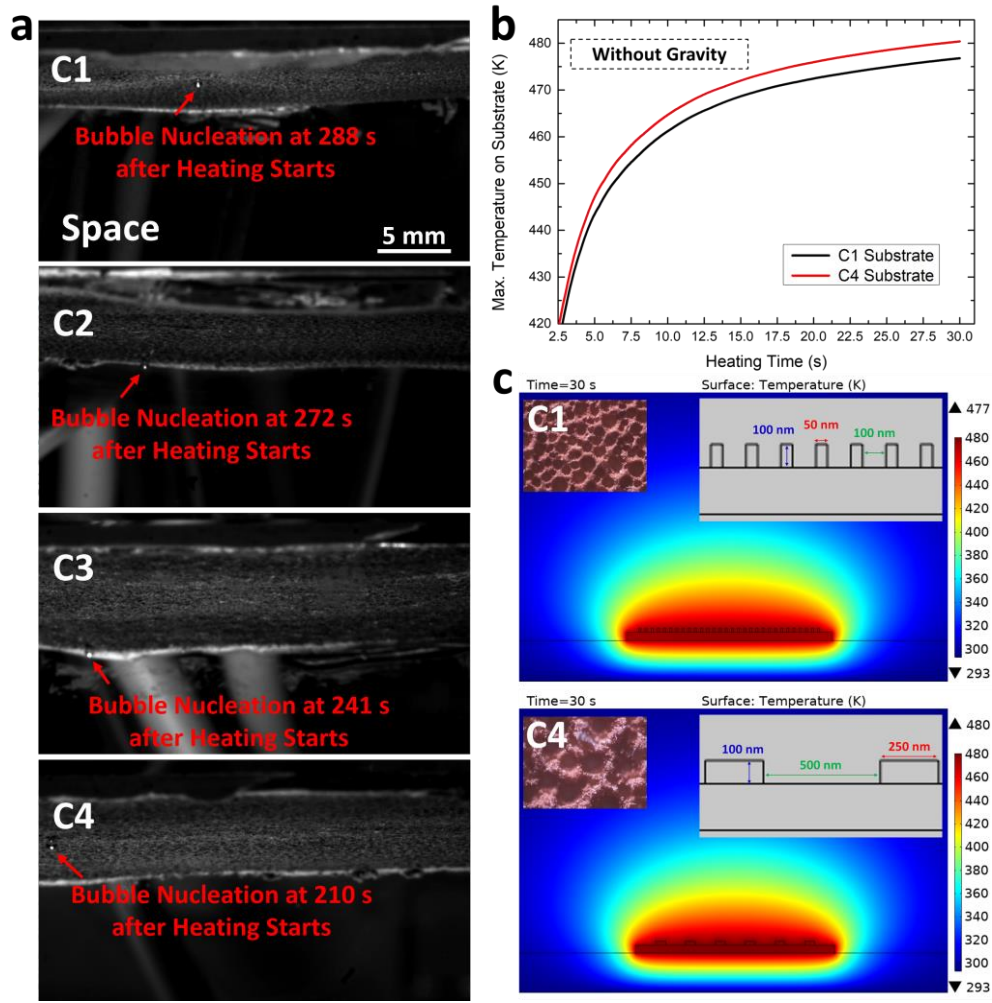

**Figure 4.** (a) The snapshots showing the surface bubble nucleation on the substrates with different characteristic lengths. (b) The simulated maximum substrate surface temperatures as a function of heating time on C1 (black) and C4 (red) substrates. (c) The simulated temperature profiles at t = 30 s on C1 (upper) and C4 (lower) substrates in space. Inserts show the characterization image and dimensions of the surface microstructures.

In this section, we will move to discuss the different bubble growth behaviors in the terrestrial and space models seen in **Figures 2c** and **d**. There are two major stages in surface bubble growth.[12] Stage I is an explosive growth due to the vaporization of the liquid surrounding the nucleation site on substrate, and stage II is a slower growth phase due to the expelling of dissolved gas from the liquid surrounding the bubble. Usually, stage I takes much shorter time (~10 ms) than stage II, with the latter usually lasting for seconds to minutes. In stage I growth, the volume of surface bubble (*V*) is proportional to the square root of time, $t^{0.5}$:[12]

$$V(t) \propto \left(\frac{P}{\rho \Lambda}\right)^{1/2} \cdot t^{1/2} \tag{1}$$

where *P* is the heating power of the boiling system, $\rho$ and $\Lambda$ are the density and latent heat of water, respectively. In stage II growth, the volume of surface bubble is proportional to time, *t*:[12]

$$V(t) = \frac{1}{3} \cdot \left(\frac{R_g T}{M_g P_\infty} \frac{C_\infty}{C_s} \left|\frac{dC_s}{dT}\right| \frac{fP}{c_w \rho}\right) \cdot t \tag{2}$$

where $R_g$ is the gas constant, *T* is the local temperature of the water surrounding bubble interface, $M_g$ is the molecular mass of air, $P_\infty$ is ambient pressure, $C_s$ is the local air solubility of the water surrounding bubble interface, $C_\infty$ is the gas saturation far away from the bubble, *f* is the heating efficiency of the boiling system, and $c_w$ is the specific heat capacity of water.

Considering that the time resolution of our camera is only ~9 ms, the periods of bubble growth that can be resolved in our videos should be mainly the stage II growth. This is also evidenced by the bubble volume growth plots in **Figure 2c** that there is no steep explosive growth period (stage I) at the very beginning of the bubble life as those reported in previous works.[12,13,18] The volume growth rate of stage II bubble is described in equation (2), in which there are three variable terms: local temperature $T$, local air solubility $C_s$ and $\left|\frac{dC_s}{dT}\right|$. The air solubility in water decreases as the temperature increases, as shown in **Figure 5a**.[72,73] As we can see, the relation between $T$ and $C_s$ is nearly linear in the experimental temperature range from room temperature (~293 K) to the boiling point (~373 K), which means $\left|\frac{dC_s}{dT}\right|$ does not change significantly in this range, leaving the only two major variables to be $T$ and $C_s$. For the terrestrial bubble, the volume is almost linear with time (**Figure 2c** (black) and refs.[12,13,18]), suggesting that the volume growth rate is constant. This means the local temperature in the water boundary layer[74] around surface bubble interface should be almost constant during the stage II growth on the earth.[12] However, as we can see in **Figures 2c** and **d** (red), the bubble volume growth is nonlinear in space, i.e., the growth rate increases with heating time. Such a different behavior can be from two possibilities: the local temperature around bubble interface keeps increasing during the stage II growth in space, or space bubble growth is dominated by water vaporization (stage I growth) instead of expelling dissolved air (stage II growth). Bubble volume growth dominated by the water vaporization around bubble interface follows equation (1), which indicates the volume should be linearly proportional to $t^{0.5}$. We plotted the space bubble volume as a function of the square root of heating time, $t^{0.5}$, in **Figure 5b**. It obvious that there is not a linear relation can be found between $V$ and $t^{0.5}$, meaning that space bubble is not following the stage I growth pattern. These analyses suggest that the space bubble is also a stage II air bubble but with increasing local temperature during growth process.

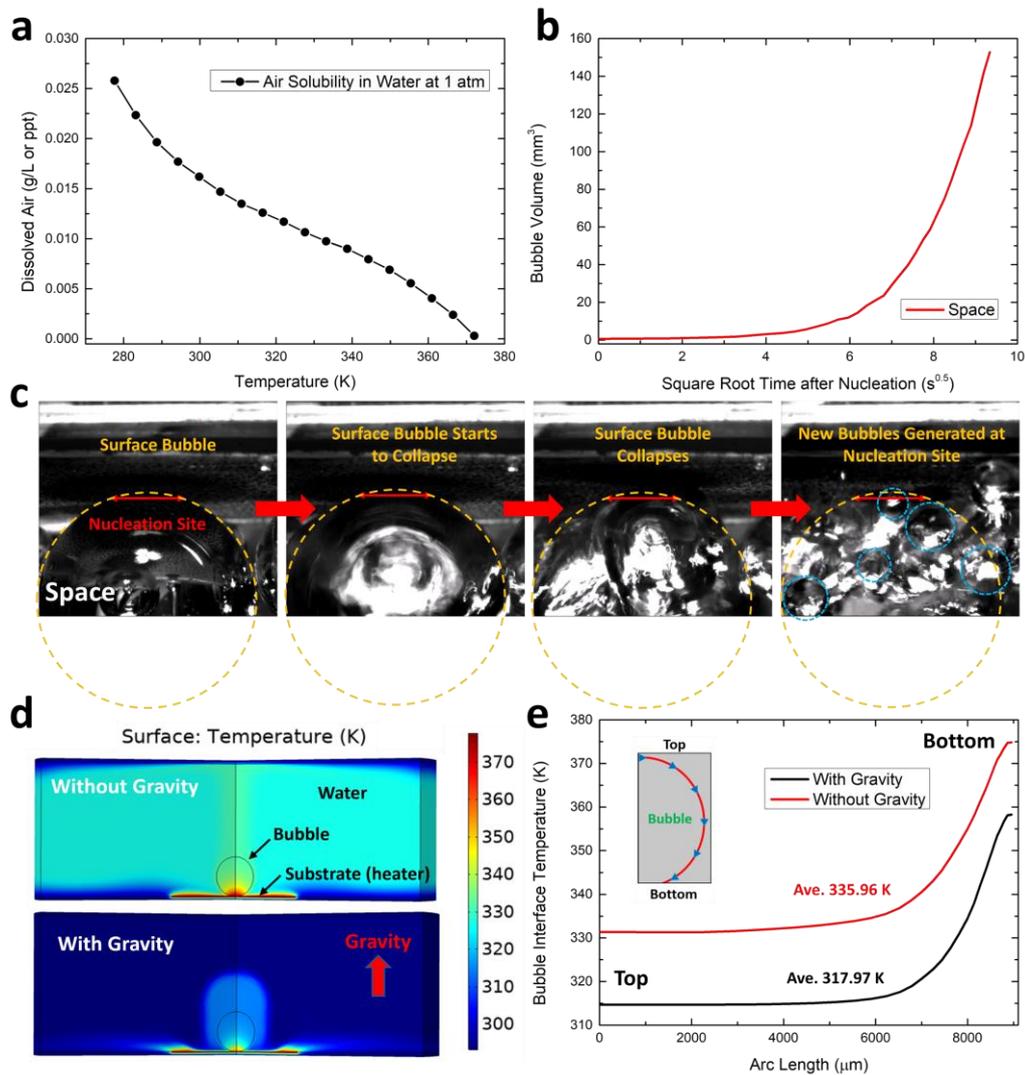

**Figure 5.** (a) Air solubility in water at 1 atm pressure as a function of temperature. (b) The space bubble volume as a function of the square root of heating time after nucleation, $t^{0.5}$. (c) The snapshots showing the collapse of a surface bubble (yellow circle) in space. After collapse, there were many smaller bubbles (blue circles) generated at the nucleation site (red line) on substrate. (d) The simulated temperature profiles around space (upper) and terrestrial (lower) bubbles. (e) The simulated bubble

interface temperature as a function of arc length in the space (red) and terrestrial (black) models. The insert shows the plotting path along bubble interface.

We know that space bubbles also mainly consist of air, but how high the local temperature around bubble interface should be in order to support such faster bubble growth (~30 times at the end of growth stage) and larger volume compared to terrestrial bubbles? Although the volume growth of air bubble is described in equation (2), the exact values of the heating efficiency ($f$), which denotes the ratio of the amount of heat used for bubble growth to the total energy input into the boiling system, and the local air solubility are unable to be measured experimentally or precisely calculated, which means we cannot directly calculate the local temperature from the measured bubble volume growth rate. However, the heating efficiency should be the same for the space and terrestrial experiments, which used the exactly same setup. As a result, the ~30 times higher bubble volume growth rate in space (**Figure 2d**) is likely contributed by the higher local temperature and its induced lower local air solubility. Looking at the air solubility-temperature relation plot in **Figure 5a**, temperature can only increase by ~1.3 times from room temperature (~293 K) to the boiling point (~373 K). That is to say, the local temperature of space bubble keeps increasing during bubble growth leading to at least the temperature around some portions of the bubble interface that near the heater surface to be close to the boiling point at the end of growth stage (before collapse). Such high local temperature can make the local air solubility decrease by over 20 times and approach to 0, which is the key to provide the extremely high bubble volume growth rate at the end of growth stage compared to the terrestrial bubble (see equation (2)). However, the high local temperature in space also involves some vaporization at bubble interface and decreases the stability of the bubble making it easy to collapse.[75] This is evidenced by the video showing the moments of space bubble collapse in **Figure 5c** (Supporting Movie S5). In addition, as we can see in the snapshots (**Figure 5c**), there were many smaller bubbles (blue circles) generated and ejected at the original nucleation site of the collapsed surface bubble. This indicates the temperature at the nucleation site was so high that vapor bubbles can be generated rapidly from it.

We conducted thermofluidic simulations to compare the bubble interface temperatures on the earth and in space using the model shown in **Figure 5d** (Supporting Information SI3). In the model, an air surface bubble with a radius of 3 mm (the size was obtained from **Figure 2c**) was added to the nucleation model that used in **Figure 3**. The contact angle of the bubble on top of the heating substrate was built according to Ref.[12] Similar to the nucleation simulations, the only difference between the terrestrial and space models is whether the gravity was considered. Based on the analysis above, we set the bubble interface temperature at the bottom of bubble around the nucleation region close to the boiling point (~373 K) in space, and then conducted the simulation of terrestrial model with the same heating power and efficiency as the space model. The simulated steady state temperature profiles of space and terrestrial bubbles are shown in **Figure 5d**. We also plotted the bubble interface temperatures from the top to the bottom of bubble along its interface in **Figure 5e**. The highest temperature is located at the bottom of bubble, which contributes the major portions of the bubble volume growth rate due to the higher temperature and its induced lower air solubility. The interface temperature of space bubble (red) is ~20 K higher than the terrestrial bubble (black). This is again because the heat transfer is much faster in the terrestrial model where convection as well as Marangoni flow[76–78] can quickly transfer the heat away from the substrate surface (see **Figure 5d**). As a result, the key issue leading to the fast bubble growth and large bubble volume in space is found to be the high local temperature and its induced low air solubility near the bubble base. Since the local temperature can be as high as the water boiling point at the end of growth stage, we believe that the ratio of vapor inside the space bubble right before collapse should be significantly higher than terrestrial bubble.

**CONCLUSION**

In summary, the nucleation and growth dynamics of surface bubbles on the earth and in space have been investigated experimentally and theoretically in this work. Due to the weak gravity field in space, the thermal convective flow is negligible compared to the case on the earth, which results in much higher local temperature around the bubble nucleation site. Such high local temperature can significantly accelerate the surface bubble nucleation and reduce the heating time required by about half. Moreover, we found the local temperature around bubble interface can be close to water boiling point and lead to extremely fast bubble growth (~30 times faster than terrestrial bubble) and large bubble volume in space. We also demonstrated that the finer the microstructures on the heating substrate, the longer the bubble nucleation time. This is mainly because the microstructure will behave as the fin structure to enhance heat conduction, and finer fin structure has higher heat conduction efficiency. These results provide fundamental insights into surface bubble dynamics, which might provide guidance on designing bubble-based sensors.[57]

## ACKNOWLEDGEMENTS

This work is supported by the Center for the Advancement of Science in Space (GA-2018-268). T. Luo would also like to thank the support from the Dorini Family endowed professorship in energy studies. We also appreciate the supports in design and performing of the experiments from Space Tango Inc. and the National Aeronautics and Space Administration (NASA), respectively.

## AUTHOR CONTRIBUTIONS

Q. Zhang, D. Mo, E. Lee and T. Luo designed the experiments, and Q. Zhang designed and performed the simulations. J. Janowitz, D. Ringle, D. Mays, A. Diddle, J. Rexroat designed and manufactured the experimental facility "CubeLab", and it was launched to the ISS twice via SpaceX Cargo Dragon 22 and Northrop Grumman. Q. Zhang and D. Mo wrote the manuscript, E. Lee and T. Luo revised it.

## COMPETING INTERESTS

The authors declare no conflict of interest.

# Supporting Information for

# Bubble Nucleation and Growth on Microstructure Surface under Microgravity

Qiushi Zhang[1], Dongchuan Mo[1], Jiya Janowitz[2], Dan Ringle[2], David Mays[2], Andrew Diddle[2], Jason Rexroat[2], Eungkyu Lee[1,*], and Tengfei Luo[1,3,*]

1. Department of Aerospace and Mechanical Engineering, University of Notre Dame, IN, USA

2. Space Tango Inc., 611 Winchester Rd. Lexington, KY, USA

3. Department of Chemical and Biomolecular Engineering, University of Notre Dame, IN, USA

* Corresponding authors: eleest@khu.ac.kr; tluo@nd.edu.

## SI1. Fabrication of Microstructured Cu Substrates.

As shown in the **Figure 1a** of main text, these microstructured Cu substrates were fabricated by the so-called hydrogen bubble template electrodeposition method.[1,2] The Cu substrates were prepared as cylinders with a diameter of ~35 mm and thickness of 0.5 mm before being cleaned with dilute sulfuric acid, hot dilute caustic solution, and deionized water.[3] A cleaned Cu substrate was then used as the cathode in the setup shown in **Figure 1a**. Another Cu plate was placed ~2 cm apart from the cathode substrate to act as the anode. The electrodeposition process was performed in a stationary solution in which the molarity of $H_2SO_4$ was kept at 0.8 M with the molarity of $CuSO_4$ ranging from 0.2 M to 1.0 M for different substrates (C1 to C4). A DC power supply (Maynuo 8852) was used for the deposition process, in which the Cu atoms in anode were dissolved into the solution and formed $Cu^{2+}$ ions. These $Cu^{2+}$ ions were driven by the electric field to move toward and finally be deposited onto the cathode Cu substrate. However, if the input current density is high enough, a hydrogen evolution reaction can occur simultaneously with the $Cu^{2+}$ ions deposition process on the cathode to initiate the hydrogen bubble template electrodeposition (**Figure 1a**). These abundant hydrogen bubbles generated on the cathode can be used as the template to construct microporous structures on the cathode Cu substrate. By controlling the molarity of $CuSO_4$, we can control the porosities of the microporous structures on Cu substrates, i.e., the porosity increases as the molarity of $CuSO_4$ increases. The deposition process lasted for 60 s each with a current density of 1 $A·cm^{-2}$. After the Cu substrates were rinsed with deionized water and dried, they were sintered in reducing atmosphere at 710 ºC for 30 mins to strengthen the microstructure.[4] **Figure 2b** shows the optical microscope images of the Cu substrates.

**SI2. Finite element thermofluidic surface bubble nucleation transient simulations.**

We employed COMSOL Multiphysics to simulate the transient temperature and flow profiles around the Cu substrate in the boiling system on ground or in space. The flow effect, thermal conduction and convection (ground model) in liquid are included in our simulations. The details of the model used in our simulations are shown in **Figure 3a**. There are several conditions that have been assumed in our simulations: (1) The liquid flow and heat transfer are both transient. Figures S1 and S2 show the flow velocity fields and temperature profiles of ground and space models at t = 3, 10 and 20 s, respectively. (2) In the liquid water, the flow is laminar (compressible with gravity on ground, compressible without gravity in space, see **Figure 3a** for the direction of gravity on ground), which satisfies the following momentum equation:

on ground,

$$\rho \frac{\partial \vec{u}}{\partial t} + \rho(\vec{u} \cdot \nabla)\vec{u} - \nabla \cdot \left( \mu(\nabla \vec{u} + \nabla \vec{u}^T) - \frac{2}{3}\mu(\nabla \cdot \vec{u})\vec{I} - p\vec{I} \right) - \rho \vec{g} = 0 \quad (s1)$$

, and in space,

$$\rho \frac{\partial \vec{u}}{\partial t} + \rho(\vec{u} \cdot \nabla)\vec{u} - \nabla \cdot \left( \mu(\nabla \vec{u} + \nabla \vec{u}^T) - \frac{2}{3}\mu(\nabla \cdot \vec{u})\vec{I} - p\vec{I} \right) = 0 \quad (s2)$$

and continuity equation:

$$\frac{\partial \rho}{\partial t} + \nabla \cdot (\rho \vec{u}) = 0 \quad (s3)$$

where $\rho$ is the density of water, $\mu$ is the dynamic viscosity of water, $\vec{u}$ is the velocity vector, $p$ is pressure, $t$ is time, $\vec{g}$ is gravity constant, and $\overleftrightarrow{I}$ is a 3×3 identity matrix. (3) The SiO$_2$ cuvette and Cu substrate are considered as rigid solid materials. (4) The heat generation rate ($Q$) of Cu substrate is the only heat source, which supplies the heat to the liquid water with the following heat transfer equations:

In water,

$$\rho C_p \frac{\partial T}{\partial t} + \rho C_p \vec{u} \cdot \nabla T - k_w \nabla^2 T = Q \tag{s4}$$

where $C_p$ is the heat capacity of water at constant pressure, $T$ is the temperature, $k_w$ is the thermal conductivity of water, $Q$ is the heat generation rate by Cu substrate, and in the medium of SiO$_2$ or Cu,

$$-k_s \nabla T = q \tag{s5}$$

where $k_s$ is the thermal conductivity of SiO$_2$ or Cu, and $q$ is the heat flux coming through the liquid/solid interfaces. The boundary conditions used in our simulations are similar to those in ref. [5–7]. The heat generation rate is:

$$Q = \frac{P}{V_0} \tag{s6}$$

where $P$ is the heating power and $V_0$ is the volume of the Cu substrate. To note, the heat generation rate of space models (C1 and C4 substrates) and ground model is the same, which is calibrated by letting the max. temperature on the substrate on ground being slightly higher than the nucleation temperature (422 K) at the end of simulation (**Figure**

**3b**).

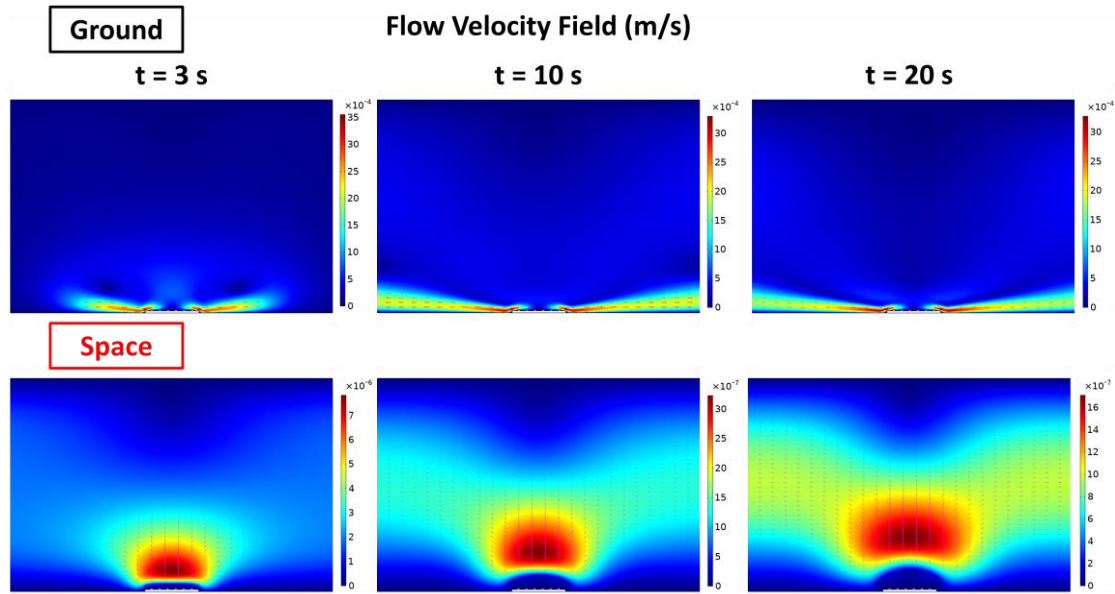

**Figure S1.** The flow velocity fields of ground (upper) and space (lower) models at t = 3, 10 and 20 s.

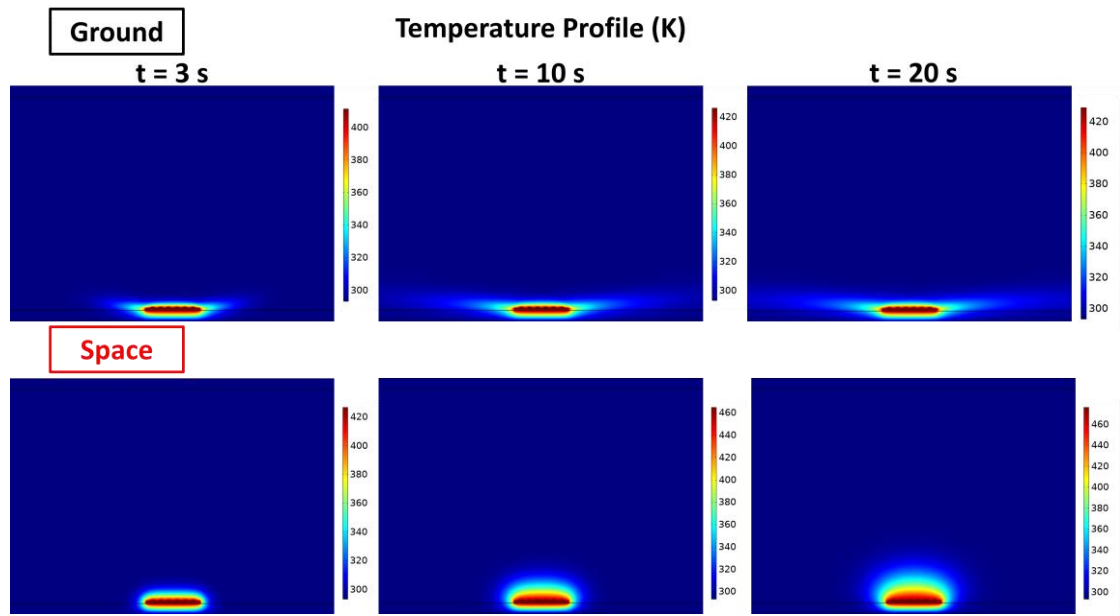

**Figure S2.** The temperature profiles of ground (upper) and space (lower) models at t = 3, 10 and 20 s.

As the experimental setup described in **Figure 1c** shows, the heat was provided by a Peltier (10 mm × 10 mm) affixed to the outside of the cuvette and conducted through the cuvette wall to the Cu substrate (20 mm × 20 mm) for surface bubble nucleation to occur. However, in order to simply our simulation models, we instead set the Cu substrate as the heat source of the boiling system in this work. To check if the heating effect of our model can represent the real experimental setup, we repeated the simulations of bubble nucleation on ground and in space while using a more realistic model, i.e., with another layer of Cu (half of the size of the Cu substrate) as the external heater to heat up the Cu substrate in the boiling system (Figure S3a). As we can see in the simulated temperature profile (Figure S3a), the heat from the external heater is conducted through the cuvette wall and mostly concentrated on the Cu substrate without any significant leakage into the cuvette wall that not covered by the Cu substrate. This is because the heat conductivity of Cu is ~2 orders of magnitude larger than $SiO_2$. As a result, this indicates that by setting the Cu substrate as the heater of the boiling system can achieve similar heating effect as the experimental system, which is also evidenced by the similar flow field and max. surface temperature plots shown in Figures S3b and c, respectively (compared to **Figures 3b and c**).

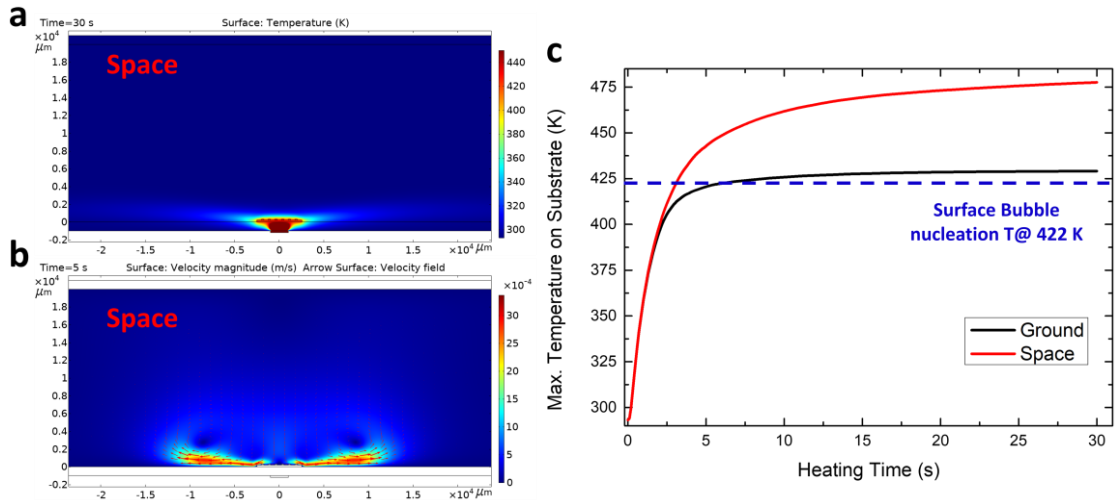

**Figure S3.** The simulated temperature profile at t = 30 s (a) and flow velocity field at t = 5 s (b) of space model using the model with external heater (compared to **Figures 3b and c**). (c) The simulated max. temperatures on substrate as a function of heating time on ground (black) and in space (red) using the model with external heater. The bubble nucleation temperature at ~422 K is indicated by a blue dash line.

**SI3. Finite element thermofluidic surface bubble growth steady-state simulations.**

The model and assumptions of bubble growth simulations are similar to those in the bubble nucleation simulations described in section SI2, while the liquid flow and heat transfer are at steady state. In liquid water, the momentum equation is:

on ground,

$$\rho(\vec{u} \cdot \nabla)\vec{u} - \nabla \cdot (\mu(\overleftrightarrow{\nabla u} + \overleftrightarrow{\nabla u^T}) - p\overleftrightarrow{I}) - \rho\vec{g} = 0 \quad \text{(s7)}$$

, and in space,

$$\rho(\vec{u} \cdot \nabla)\vec{u} - \nabla \cdot (\mu(\overleftrightarrow{\nabla u} + \overleftrightarrow{\nabla u^T}) - p\overleftrightarrow{I}) = 0 \quad \text{(s8)}$$

, and continuity equation is:

$$\rho(\nabla \cdot \vec{u}) = 0 \quad \text{(s9)}$$

where the definitions of variables are the same as those in equations (s1), (s2) and (s3), respectively. The heat transfer equation in water is:

$$\rho C_p \vec{u} \cdot \nabla T - k_w \nabla^2 T = Q \quad \text{(s10)}$$

where the definitions of variables are the same as those in equation (s4). The heat generation rate also follows the same equation as equation (s6). An air bubble with a radius of 3 mm was added on top of the Cu substrate (**Figure 5d**). The gas medium inside the surface bubble, quartz walls and Cu substrate were considered as non-fluidic rigid materials, which have the same heat transfer equation as equation (s5). The interface of the bubble (gas/water boundary) has a slip boundary condition with the Marangoni effect as:

$$\left[\mu(\overleftrightarrow{\nabla u} + \overleftrightarrow{\nabla u}^T) - \left(p + \frac{2}{3}\mu(\nabla \cdot \vec{u})\right)\overleftrightarrow{I}\right]\hat{n} = \gamma \nabla_t T \qquad (s11)$$

where $\hat{n}$ is the normal outward vector to the surface of the bubble, $\gamma$ is the temperature derivative of the water/gas surface tension, and $\nabla_t$ is the gradient of the tangent vector to the surface of the bubble.